# Analysis of chaotic flow in a 2D multi-turn closed-loop pulsating heat pipe

S. M. Pouryoussefi[1] and Yuwen Zhang[1*]

[1]*Department of Mechanical and Aerospace Engineering, University of Missouri, Columbia, MO, 65211, USA*

**Abstract**

Numerical study has been conducted for the chaotic flow in a multi-turn closed-loop pulsating heat pipe (PHP). Heat flux and constant temperature boundary conditions have been applied for heating and cooling sections respectively. Water was used as working fluid. Volume of Fluid (VOF) method has been employed for two-phase flow simulation. Volume fraction results showed formation of perfect vapor and liquid plugs in the fluid flow of PHP. Non-linear time series analysis, power spectrum density, correlation dimension and autocorrelation function were used to investigate the chaos. Absence of dominating peaks in the power spectrum density was a signature of chaos in the pulsating heat pipe. It was found that by increasing the filling ratio and evaporator heating power the correlation dimension increases. Decreasing of the autocorrelation function with respect to time showed the prediction ability is finite as a result of chaotic state. An optimal filling ratio of 60% and minimum thermal resistance of 1.62 °C/W were found for better thermal performance of the pulsating heat pipe.

## 1. INTRODUCTION

Downsizing of personal computers and advancing performance of processors has called for the development of micro and miniature heat pipes to transport heat from chips to heat sinks. The Pulsating or Oscillating Heat Pipe (OPH or PHP) is a very promising heat transfer device. Due to the pulsation of the working fluid in the axial direction of the tube, heat is transported from the evaporator section to the condenser section. The heat input, which is the driving force, increases the pressure of the vapor plug in the evaporator section. In turn, this increased pressure will push the neighboring vapor plugs and liquid slugs toward the condenser, which is at a lower pressure [1]. Although a variety of designs are in use, the fundamental processes and parameters affecting the PHP operation still need more investigation. Shafii et al. [2] presented analytical models for both open- and closed-loop PHPs with multiple liquid slugs and vapor plugs. Heat transfer in both looped and unlooped PHPs was due mainly to the exchange of sensible heat and higher surface tension resulted in a slight increase in total heat transfer.

Zhang and Faghri [3-5] investigated heat transfer process in evaporator and condenser sections of the PHP. They developed heat transfer models in the evaporator and condenser sections of a pulsating heat pipe with one open-end by analyzing thin film evaporation and condensation. Results showed the frequency and amplitude of the oscillation is almost unaffected by the surface tension after steady oscillation has been established. The amplitude of oscillation was decreased with decreasing diameter of the pulsating heat pipe and decreasing wall temperature of the heating section, but the frequency of oscillation was almost unchanged.

Researchers have conducted many experimental and theoretical studies to investigate complicated behaviors and unsolved issues of the pulsating heat pipes [6-12]. Recent studies have suggested the existence of chaos in PHPs under some operating conditions [13-16]. The approach in these studies is to analyze the time series of fluctuation of temperature of a specified location on the PHP tube wall (adiabatic section) by power spectrum calculated through Fast Fourier Transform (FFT). The two dimensional mapping of the strange attractor and the subsequent calculation of the Lyapunov exponent have been performed to prove the existence of chaos in PHP system. By calculating Lyapunov exponents, it was shown that the theoretical models are able to reflect the characteristic chaotic behavior of experimental devices.

Although several theoretical and experimental studies of the chaotic behavior of pulsating heat pipes have been carried out, there has been no detailed numerical simulation for PHPs. Pouryoussefi and Zhang [17] recently conducted a numerical simulation of the chaotic flow in the closed-loop pulsating heat pipe with two turns. Constant temperature was the boundary condition for both evaporator and condenser. Chaotic behavior of the PHP was investigated under different operating

---

*Corresponding author: zhangyu@missouri.edu, Phone: +1-573-884-6936, Fax: +1-573-884-5090

conditions. In this paper, the authors have extended their previous work [17] by applying heat flux boundary condition and thermal behavior investigation in addition to chaotic behavior investigation. Heat flux and constant temperature boundary conditions have been used for evaporator and condenser, respectively. The PHP structure is two-dimensional and water is the working fluid. Thermal resistance, axial wall temperature distribution and chaotic parameters have been investigated under different operating conditions.

## 2. PHYSICAL MODELING

The two dimensional structure of the PHP were the same for all different operating conditions in this work and water was the only working fluid. Different evaporator heating powers, condenser temperatures and filling ratios were tested for the numerical simulation. The pulsating heat pipe structure consists of three sections: heating section (evaporator), cooling section (condenser), and adiabatic section. Figure 1 illustrates a schematic configuration of the PHP which has been used in this study. The three different sections have been distinguished by two horizontal lines. The length and width of the tube are 870 mm and 3 mm respectively. The span of the PHP was considered 1 m as a 2D simulation. Figure 2 shows meshing configuration used in this study. Only a part of evaporator section has been depicted to show the quadrilaterals mesh which were employed for simulation. The quadrilaterals mesh was used for the entire pulsating heat pipe.

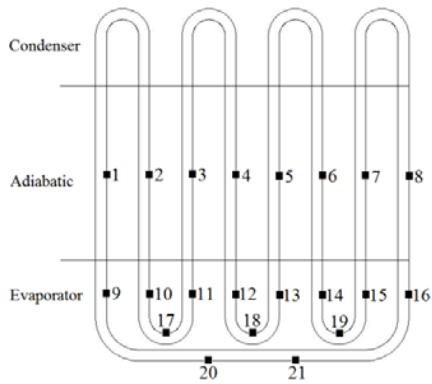

Fig. 1. Pulsating heat pipe structure.

Volume of Fluid (VOF) method has been applied for two phase flow simulation. The VOF model is a surface-tracking technique applied to a fixed Eulerian mesh. It is designed for two or more immiscible fluids where the position of the interface between the fluids is of interest. In the VOF model, a single set of momentum equations is shared by the fluids, and the volume fraction of each of the fluids in each computational cell is tracked throughout the computational domain.

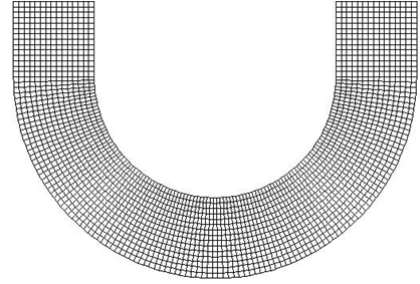

Fig. 2. Meshing configuration.

Applications of the VOF model include stratified flows, free-surface flows, filling, sloshing, the motion of large bubbles in a liquid, the motion of liquid after a dam break, the prediction of jet breakup (surface tension), and the steady or transient tracking of any liquid-gas interface. The VOF formulation relies on the fact that two or more fluids (or phases) are not interpenetrating. For each additional phase added to the model, the volume fraction of the phase in the computational cell is introduced. In each control volume, the volume fractions of all phases sum to unity. The fields for all variables and properties are shared by the phases and represent volume-averaged values, as long as the volume fraction of each of the phases is known at each location. Thus the variables and properties in any given cell are either purely representative of one of the phases, or representative of a mixture of the phases, depending on the volume fraction values. Then the $q^{th}$ fluid's volume fraction in the cell is denoted as $\alpha_q$. Based on the local value of $\alpha_q$, the appropriate properties and variables will be assigned to each control volume within the domain. Tracking of the interfaces between the phases is accomplished by the solution of a continuity equation for the volume fraction of one (or more) of the phases. For the $q^{th}$ phase, this equation has the following form [17]:

$$\frac{1}{\rho_q}\left[\frac{\partial}{\partial t}(\alpha_q \rho_q) + \nabla . (\alpha_q \rho_q \boldsymbol{v}) = S_{\alpha q} + \sum_{p=1}^{n}(\dot{m}_{pq} - \dot{m}_{qp})\right] \quad (1)$$

where $\dot{m}_{qp}$ is the mass transfer from phase q to phase p and $\dot{m}_{pq}$ is the mass transfer from phase p to phase q. $S_{\alpha q}$ is the source term on the right-hand side of equation and equal to zero. The primary-phase volume fraction will be computed based on the following constraint:

$$\sum_{q=1}^{n} \alpha_q = 1 \quad (2)$$

The volume fraction equation was solved through explicit time discretization. In the explicit approach, finite-difference interpolation schemes are applied to the volume fractions that were computed at the previous time step.

$$\frac{\alpha_q^{n+1} \rho_q^{n+1} - \alpha_q^n \rho_q^n}{\Delta t} V + \sum_f (\rho_q U_f^n \alpha_{q,f}^n) = \left[\sum_{p=1}^{n}(\dot{m}_{pq} - \dot{m}_{qp}) + S_{\alpha_q}\right] V \quad (3)$$

where n+1 is the index for new (current) time step, n is the index for previous time step, $\alpha_{q,f}$ is the face value of the $q^{th}$ volume fraction, V is the volume of cell, and $U_f$ is the volume flux through the face based on normal velocity. The properties appearing in the transport equations are determined by the presence of the component phases in each control volume. In the vapor-liquid two-phase system, the density and viscosity in each cell are given by

$$\rho = \alpha_v \rho_v + (1 - \alpha_v)\rho_l \quad (4)$$
$$\mu = \alpha_v \mu_v + (1 - \alpha_v)\mu_l \quad (5)$$

A single momentum equation is solved throughout the domain, and the resulting velocity field is shared among both phases. The momentum equation, shown below, is dependent on the volume fractions of all phases through the properties $\rho$ and $\mu$.

$$\frac{\partial}{\partial t}(\rho \vec{v}) + \nabla \cdot (\rho \boldsymbol{vv}) = -\nabla p + \nabla \cdot [\mu(\nabla \boldsymbol{v} + \nabla \boldsymbol{v}^T)] + \rho \boldsymbol{g} + \boldsymbol{F} \quad (6)$$

One limitation of the shared-fields approximation is that in cases where large velocity differences exist between the phases, the accuracy of the velocities computed near the interface can be adversely affected. The energy equation, also shared among the phases, is shown below.

$$\frac{\partial}{\partial t}(\rho E) + \nabla \cdot (\boldsymbol{v}(\rho E + p)) = \nabla \cdot (k_{eff} \nabla T) + S_h \quad (7)$$

The VOF model treats energy, E, and temperature, T, as mass-averaged variables:

$$E = \frac{\sum_{q=1}^{n} \alpha_q \rho_q E_q}{\sum_{q=1}^{n} \alpha_q \rho_q} \quad (8)$$

where $E_q$ for each phase is based on the specific heat of that phase and the shared temperature. The properties $\rho$ and $k_{eff}$ (effective thermal conductivity) are shared by the phases and the source term, $S_h$ is equal to zero.

## 3. RESULTS and DISCUSSION

### 3.1 Volume Fractions

Figure 3 illustrates volume fractions of liquid and vapor at different times (Red color represents the vapor and Blue color represents the liquid). Figure 3(a) shows almost the initial condition of the PHP. Figure 3(b) explains formation of vapor bubbles and fluid flow development in the PHP. Figure 3(c) depicts the volume fractions of liquid and vapor in the PHP after the fluid flow has been established. Mostly a similar process occurs in the PHP with filling ratio of 60% (Fig. 4). One of the most significant effects due to the increasing the filling ratio is pressure increment in the PHP which lead to a longer time duration for boiling start (because of increasing the saturated temperature) and slower flow motion in the PHP. The effect of change in filling ratio on the chaotic and thermal behavior of the PHP will be investigated in the next sections afterwards. It is evident liquid plugs having menisci on the plug edges are formed due to surface tension forces. A liquid thin film also exists surrounding the vapor plug. The angle of contact of the menisci, the liquid thin film stability and its thickness depends on the fluid-solid combination and the operating parameters which are selected. If a liquid plug is moving or tends to move in a specific direction then the leading contact angle (advancing) and the lagging contact angle (receding) may be different [13]. This happens because the leading edge of the plug moves on a dry surface (depending on the liquid thin film stability and existence) while the lagging edge moves on the just wetted surface. These characteristics can be observed in Fig. 5 as result of numerical simulation.

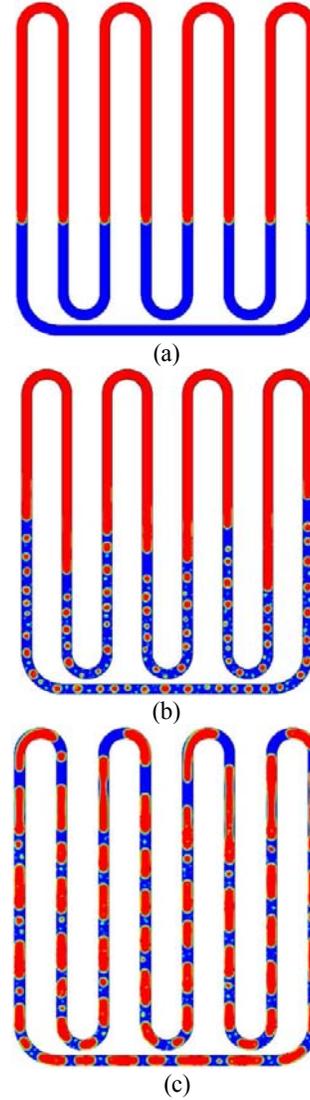

Fig. 3. Volume fractions of liquid and vapor at different times with Q=40 W, $T_c$=30 °C and FR= 35%; (a): t=0.1 s, (b): t=1.7 s, (c): t=20 s.

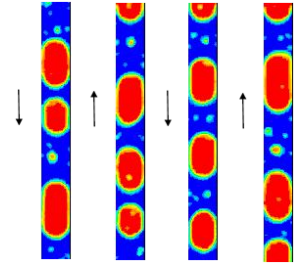

Fig. 5. Vapor and liquid plugs in the PHP

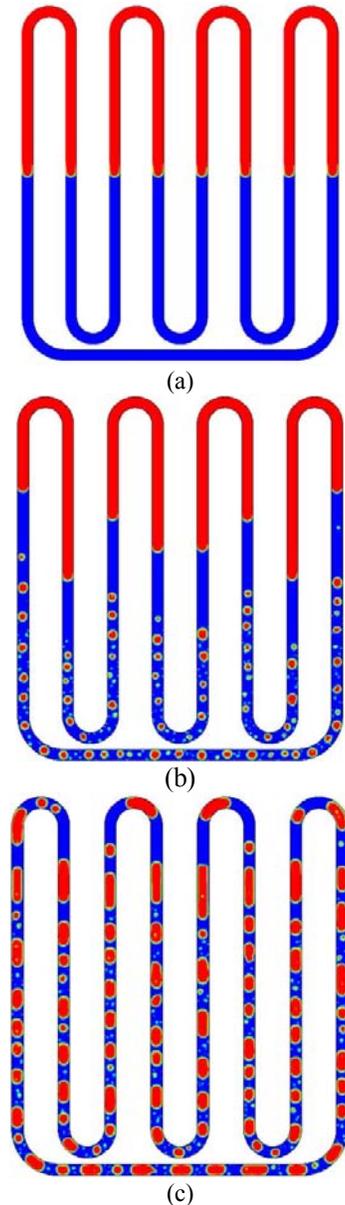

Fig. 4. Volume fractions of liquid and vapor at different times with Q=55 W, $T_c$=25 °C and FR= 60%; (a): t=0.1 s, (b): t=2.6 s, (c): t=20 s.

The results showed that the fluid flow finally circulates in one direction (clockwise or counterclockwise) in the pulsating heat pipe. This direction is based on a random process and could be different even under the same operating and boundary conditions [17]. It is seen that by increasing the wall temperature, the liquid film thickness around the vapor plugs decreases. In evaporator section the temperature of the liquid plugs increases which is followed by evaporation mass transfer to the adjoining vapor plug or splitting the liquid plug by formation of new bubbles inside due to the nucleate boiling in the slug flow regime. Sometimes two vapor bubbles combine together to form a larger vapor slug. This phenomenon was mostly observed in the adiabatic section and rarely was observed in the condenser. In addition, liquid accumulation as a result of vapor condensation occurred on the condenser surface. This condensation process was more visible on the bending parts due to the surface tension. It was observed that many of the simulation results failed to get a successful performance for the pulsating heat pipe because of the operating condition. In some cases, the fluid was stuck in the condenser section due to the low filling ratio and low temperature of the condenser section which leads to higher surface tension for that part. In some cases, high temperature of the evaporator surface led to form abnormal vapor plugs in this section due to the intensive nucleate boiling and formation of superheated vapor.

Figure 6 compares formation of such abnormal vapor plugs with normal plugs. Formation of abnormal vapor plugs had adverse effect on the fluid flow in the PHP which decreases performance of the PHP. After observing results of simulations for volume fraction contours and temperature time series, many of them were discarded and those that seem to have proper behavior used for further investigation of chaotic flow.

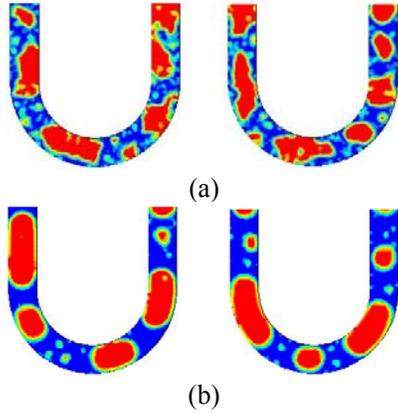

Fig. 6. Comparing formation of abnormal vapor plugs (a) with normal vapor plugs (b) at evaporator

## 3.2 Non-linear temperature oscillations

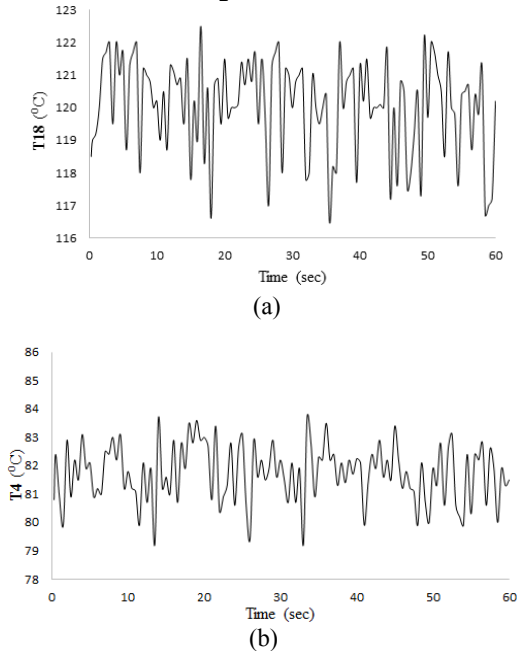

Fig. 7. Temperature oscillations at evaporator (a) and adiabatic wall (b)

The non-linear time series analysis is a popular method for investigation of complicated dynamical systems. Different non-linear analytical approaches, including power spectrum density, correlation dimensions, and autocorrelation function have been employed to analyze the temperature time series at adiabatic wall. Temperature behavior of adiabatic wall in a PHP is one of the most important factors to investigate chaos in the PHP. It can be seen in Fig. 1 that there are 21 points on the wall of the PHP assigned for temperature measurements. Eight points are located at the adiabatic section and thirteen points are located at evaporator. Then, there are eight choices to observe temperature behavior of adiabatic wall. Figure 7 illustrates the temperature oscillations of point #4 and #18 on the adiabatic wall and evaporator respectively on the PHP under operating condition mentioned in Fig. 4. Point selection at adiabatic section and evaporator was done randomly.

It is obvious that aperiodic, irregular, non-linear and complex time series of temperature are obtained for both evaporator and adiabatic wall. Evaporator temperature oscillations have higher amplitude and higher mean temperature comparing with adiabatic wall temperature oscillations. As mentioned earlier only adiabatic wall temperature behavior has been considered for chaos investigation purpose. To reach this aim, power spectrum density analysis has been applied to the obtained time series of adiabatic wall temperature. The PSD is the average of the Fourier transform magnitude squared over a large time interval. Then it can be defined as [18]:

$$S_x(f) = \lim_{T \to \infty} E \left\{ \frac{1}{2T} \left| \int_{-T}^{T} x(t) e^{-j2\pi ft} dt \right|^2 \right\} \quad (9)$$

where x(t) is the random time signal and E is the energy of the signal.

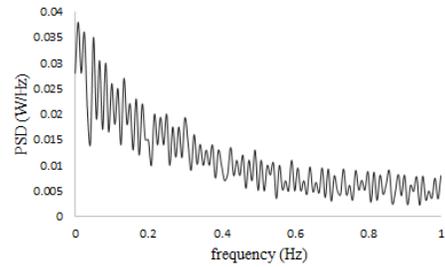

Fig. 8. Power spectrum density of the adiabatic wall temperature time series (point #4)

Figure 8 shows the power spectrum density of the time series for the temperature on the adiabatic wall by Matlab analysis. Absence of dominating peaks in power spectrum density behavior and its decay by increasing the frequency were signatures of chaotic state in the PHP under such operating conditions. Important approaches such correlation dimension and autocorrelation function were employed to investigate the chaotic state afterwards. More details of the calculations in the mentioned approaches can be found in the references [17, 18].

## 3.3 Correlation dimension and Autocorrelation function

Finding the correlation dimension for a chaotic process from the time series data set is usually applied for getting information about the nature of the dynamical system. Correlation dimension is popular because of its simplicity and flexibility data storage requirements. It is an alternate definition for true dimension or fractal dimension [17]. Time series data set can reconstruct phase space formed by the trajectories of the system, the attractor. Suppose that our attractor consists of N data points obtained numerically or experimentally, the correlation sum is defined as number of pairs of points within a sphere of radius r divided by the number of data points square. The correlation sum typically scales as [18]:

$C(r) \approx \alpha \, r^{Dc}$  (10)

where $D_c$ is the correlation dimension and r is chosen to be smaller than the size of attractor and larger than the smallest spacing between the points. In practice, we plot ln C(r) as a function of ln r and obtain the dimension $D_c$ from the slope of the curve.

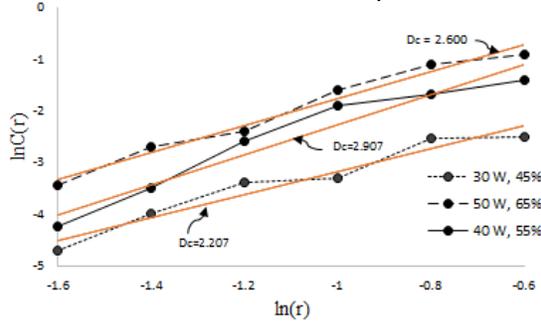

Fig. 9. Correlation dimension values ($D_c$)

Figure 9 illustrates the correlation dimension for evaporator heating powers of 30 W, 40 W and 50 W with filling ratios of 45%, 55% and 65%, respectively at condenser temperature of 35 °C. Values of 2.207, 2.600 and 2.907 were obtained for correlation dimension. It is evident that by increasing the evaporator heating power and filling ratio, the slope of the curves and correlation dimension increase. High values of correlation dimension refers to high frequency, small scale temperature oscillations, caused by miniature bubbles or short vapor plugs dynamically flowing in PHP tubes [17]. The lower correlation dimension corresponds to low frequency of temperature oscillations and large amplitude caused by large bubbles in the PHP.

The theoretical autocorrelation function is an important tool to describe the properties of a stochastic process. An important guide to the persistence in a time series is given by the series of quantities called the sample autocorrelation coefficients, which measure the correlation between observations at different times. The set of autocorrelation coefficients arranged as a function of separation in time is the sample autocorrelation function, or the ACF. AFC helps to determine how quickly signals or process change with respect to time and whether a process has a periodic component. Figure 10 illustrates the autocorrelation function (AFC) of time series at filling ratio of 65% and three evaporator heating powers of 30 W, 45 W and 60 W at condenser temperature of 35 °C. Autocorrelation function of the time series are computed as [18]:

$$ACF(\tau) = \frac{\sum_{i=1}^{N}(T_{i+\tau}-\bar{T})(T_i-\bar{T})}{\sum_{i=1}^{N}(T_i-\bar{T})^2}$$  (11)

where $T_i$ and $T_{i+\tau}$ are the adiabatic wall temperature observations at the time domain, $\bar{T}$ is the overall mean temperature and τ is the lag.

Behavior of ACF indicates the prediction ability of the system [17]. The autocorrelation function will have its largest value of AFC=1 at τ=0. Figure 10 shows that the autocorrelation function decreases with time. Decreasing of the ACF shows that the prediction ability is finite as a signature of chaos. It is obvious in Fig. 10 by increasing the evaporator heating power, ACF decreases.

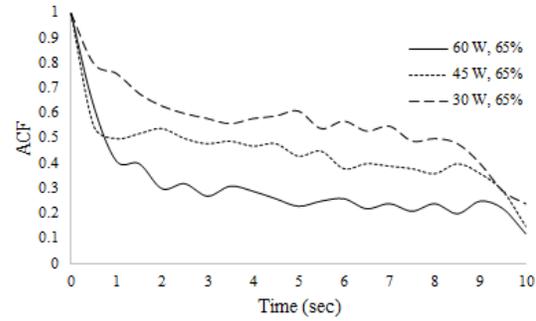

Fig. 10. Autocorrelation function versus time

### 3.4 Thermal behavior

For the traditional heat pipes, the filling ratio range of working fluid is lower comparing to the PHPs. Large amount of working fluid inside the channels causes liquid blockage and reduces pressure difference between heating and cooling sections which can stop the oscillation movement. But for the PHPs a higher filling ratio is required than traditional heat pipes. To investigate the effect of filling ration on the performance of PHP, several simulations were carried out under different heating powers, condenser temperatures and filling ratios. Figures 11(a) and 11(b) explain the axial temperature of the wall through the PHP. It should be noted that the relative distance for the point #1 is considered zero as the origin and the distance increases by moving downward from point #1 to #9. In addition, the temperature distributions in these figures do not include the condenser temperature since its value is known and constant. Temperature profiles have been illustrated at three different filling ratios of 40%, 60% and 80% under heating powers of 40 W and 55 W for evaporator and condenser temperature of 25 ˚C as part of numerical simulations. Mean temperature for each distance is defined as the average of data points from 0 to 60 seconds on that distance.

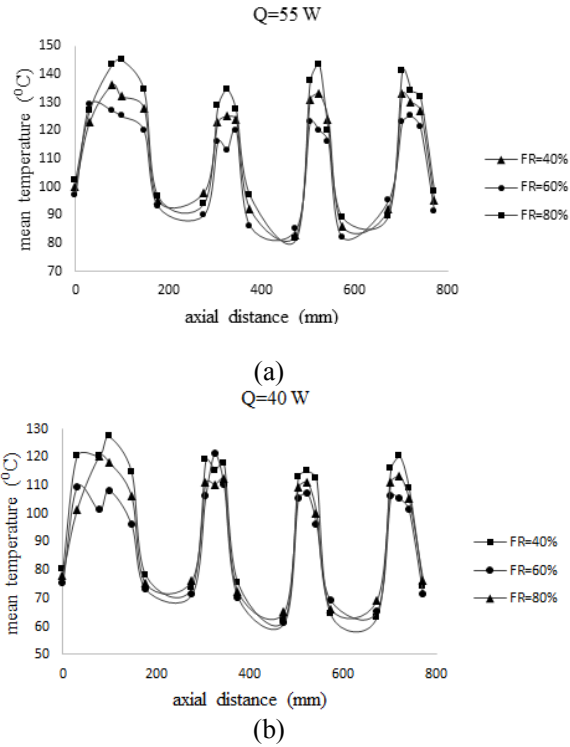

(a)

(b)

Fig. 11. Axial mean temperature distribution for heating powers of 55 W (a) and 40 W (b)

Generally these temperature distributions were symmetric from point #2 to #3, #4 to #5, #6 to #7 and #20 to #21 for vertical centerlines at #17, #18, #19 and #18 respectively. It was found that for all the heating powers, the mean temperatures in the evaporator had its lowest values at the filling ratio of 60%. It is obvious that at higher filling ratio of 80% and lower filling ratio of 40%, the evaporator mean temperatures have higher values comparing to that of 60%. Then based on the numerical simulation results, filling ratio of 60% was found as optimal filling ratio for the thermal performance of the PHP.

The overall thermal resistance of a PHP is defined as the difference average temperatures between evaporator and condenser divided by the heating power. For evaporator average temperature, thirteen points from #9 to #21 were used to calculate the average temperature. Since condenser has constant temperature, calculation is not necessary to get the average temperature which is equal to that constant value. Equation (12) defines the thermal resistance of a PHP.

$$R = \frac{T_e - T_c}{Q} \qquad (12)$$

where $T_e$ is the evaporator average temperature, $T_c$ is the condenser average temperature and Q is the evaporator heating power.

Figure 12 describes behavior of thermal resistance with respect to evaporator heating power at recommended optimal value of 60% for filling ratio and condenser temperature of 25 ˚C. By increasing the heating power from 10 to 40 W, the thermal resistance decreases sharply. From 40 to 45 W, thermal resistance has slight decrease and remains almost constant about 1.62 ˚C/W at heating power of 45 W. Then thermal resistance starts to increase by increasing the heating power afterwards. It was seen that the PHP has its lowest thermal resistance of 1.62 ˚C/W at evaporator heating power of 45 W. It should be noted that the thermal resistance depends on many parameters such as PHP structure, working fluid and operating conditions. Thus, obtained values in this paper are suitable for the test case in current study. But it can be concluded that there are optimal filling ratios and minimum thermal resistance at any test case which can be obtained by numerical or experimental investigations.

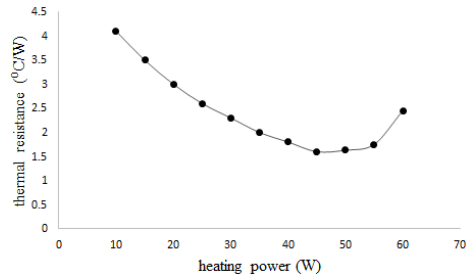

Fig. 12. Thermal resistance versus heating power

## 4. CONCLUSIONS

Numerical simulations have been implemented to investigate the chaotic and thermal behavior of a closed-loop pulsating heat pipe. Heat flux and constant temperature boundary conditions were applied for evaporator and condenser respectively. Investigations have been conducted over several operating conditions. Volume fraction results showed formation of perfect vapor and liquid plugs in the fluid flow of PHP. Non-linear time series analysis, Power spectrum density, Correlation dimension and Autocorrelation were used to investigate chaos. Chaotic behavior was observed under several operating conditions. An optimal filling ratio and minimum thermal resistance were found for better thermal performance of the pulsating heat pipe.

## NOMENCLATURE

$\alpha_q$ : Void Fraction of phase q
$v_q$ : Velocity of phase q (m/s)
$\rho_q$ : Density of phase q (kg/m$^3$)
$\mu$ : Dynamic viscosity (kg/m.s)
$k_{eff}$ : Effective thermal Conductivity (W/m.K)
$g$ : Gravity acceleration (m/s$^2$)
$f$ : Frequency (Hz)